# Quantum Mechanical Treatment of Stimulated Raman Cross Sections

Wei Min[1] [*] and Xin Gao[1]

[1] Department of Chemistry, Columbia University, New York, NY 10027, USA

[*]Corresponding author: wm2256@columbia.edu (W.M.)

## Abstract

Stimulated Raman scattering (SRS) has played an increasingly pivotal role in chemistry and photonics. Recently, understanding of light-molecule interaction during SRS was brought to a new quantitative level through the introduction of stimulated Raman cross section, $\sigma_{SRS}$. Measurements of Raman-active molecules have revealed interesting insights, and theoretical consideration has suggested an Einstein-coefficient-like relation between $\sigma_{SRS}$ and the commonly used spontaneous Raman cross sections, $\sigma_{Raman}$. However, the theoretical underpinning of $\sigma_{SRS}$ is not known. Herein we provide a full quantum mechanical treatment for $\sigma_{SRS}$, via both a semi-classical method and a quantum electrodynamic (QED) method. The resulting formula provides a rigorous theory to predict experimental outcome from first principles, and unveils key physical factors rendering $\sigma_{SRS}$ inherently strong response. Through this formula, we also confirm the validity of the Einstein-coefficient-like equation connecting $\sigma_{Raman}$ and $\sigma_{SRS}$ reported earlier, and discuss the inherent symmetry between all spontaneous and stimulated optical processes. Hence the present treatment shall deepen the fundamental understanding of the molecular response during SRS, and facilitate quantitative applications in various experiments.

## Introduction

Light-matter interactions are of central importance to science and technology. The processes of absorption and emission of single photons, including stimulated absorption, spontaneous emission and stimulated emission, are among the simplest and the most fundamental interactions, thanks to Einstein's insightful treatment. His A and B coefficients relate the spontaneous emission rate to the stimulated emission coefficient by requiring consistency with Planck's blackbody radiation law together with microscopic reversibility. This relation was worked out in 1917, without reference to the exact physics of the coupling of radiation to matter[1]. With the advent of quantum theory, Einstein's B coefficients for absorption and stimulated emission can be calculated by semiclassical theory, where the matter is treated quantum mechanically but the electromagnetic field is treated classically[2]. Later, in quantum electrodynamics (QED), where both the matter and the electromagnetic field are quantized, spontaneous emission is the consequence of the coupling between the excited molecule and the vacuum field fluctuation. The rate of this process, Einstein's A coefficient, can then be calculated from first principles[3].

Raman scattering, a nonlinear two-photon process, also takes these two distinct forms. Spontaneous Raman scattering was theoretically predicted 100 years ago by Smekal[4] and observed a few years later by Raman and Krishnan[5]. The effect of stimulated Raman scattering (SRS) was discovered accidentally in 1962[6], and was harnessed by SRS spectroscopy and microscopy in recent decades with broad impact in chemistry and photonics[7–11]. The strength of Raman scattering by molecules in the literature is almost exclusively characterized by Raman cross section, $\sigma_{Raman}$, which exhibits a dimension of area, even in the context of coherent Raman experiments [12–19]. Recently, stimulated Raman cross section, $\sigma_{SRS}$, was introduced phenomenologically to characterize *intrinsic* molecular response during SRS, in a similar sprit to Einstein's B coefficient capturing the response of matter during stimulated emission. It was proposed after making an analogy to two-photon absorption cross section (which has a dimension of $cm^4 \cdot s$, named after its developer Göppert-Mayer)[20]:

$$R_{SRS} = \sigma_{SRS} \cdot \phi_p \cdot \phi_S \qquad (1)$$

The measurement results of $\sigma_{SRS}$ for a series of Raman active molecules have revealed interesting insights about Raman response. Different from the prevailing view that $\sigma_{Raman}$ is always many orders of magnitude (up to $10^{14}$) smaller than its electronic absorption counterpart, $\sigma_{SRS}$ can even be much larger than the two-photon absorption cross section of similar molecules[20]. In a recent attempt to make theoretical connection between $\sigma_{Raman}$ and $\sigma_{SRS}$, a relation was derived using the concept of virtual vacuum photons[21]. The resulting equation resembles Einstein's coefficients connecting spontaneous emission and stimulated emission, and it has found utility in predicting and explaining absolute signal of SRS microscopy[22]. Also similar to the original derivation by Einstein, a physical constraint was adopted there *without* explicitly referring to full quantum mechanics, and, consequently, the theoretical expression of $\sigma_{SRS}$ was not given[21].

Herein we provide both a semiclassical treatment and a full QED treatment for $\sigma_{SRS}$. The semiclassical approach is justified, as the electromagnetic fields used in SRS experiments, especially in SRS microscopy, are macroscopic so that the photon number can be treated as a continuous variable. Additionally, we also employ QED where the light field is treated as quantized photons. Our key formula, Eq. (24), reveals the nature of $\sigma_{SRS}$ from first principles. We then show that the numerical estimates of $\sigma_{SRS}$ agree well with the experimental measurements of model compounds and elucidate key physical factors rendering $\sigma_{SRS}$ inherently strong response (up to 500,000 Göppert-Mayer). Finally, we compare the expression of $\sigma_{SRS}$ with $\sigma_{Raman}$. Indeed, we can reproduce the earlier equation between these two Raman cross sections in an independent way, and further enrich it with an integral and a differential version. Furthermore, we are prompted to present a generalized form of Einstein's coefficients, Eq. (49). Hence this study deepens our understanding on molecular Raman response, puts $\sigma_{SRS}$ on firm theoretical ground and equal footing with $\sigma_{Raman}$, and completes a symmetric analogy between spontaneous emission, stimulated emission, spontaneous Raman scattering and stimulated Raman scattering.

## Results

**Semi-classical derivation of stimulated Raman cross sections**

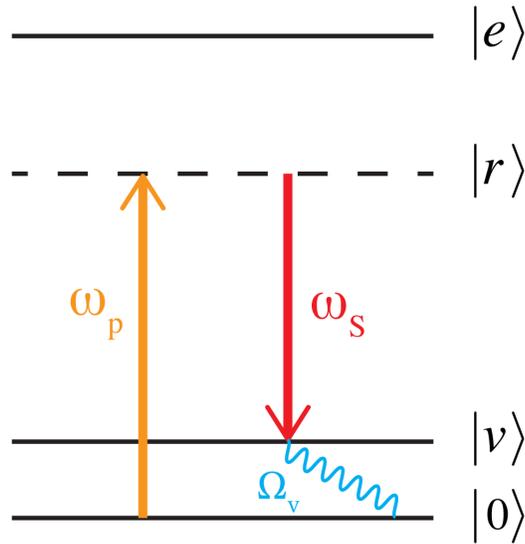

**Figure 1 Energy diagram of stimulated Raman scattering.** A molecule in its ground state $|0\rangle$ absorbs a photon from the pump beam, transiently occupies a virtual state $|r\rangle$, and concurrently emits an identical photon to the Stokes beam before returning to a vibrational state $|v\rangle$. $|e\rangle$ denotes the electronic excited state. $\omega_p$, $\omega_S$, $\Omega_v$ represent the frequencies of the pump beam, Stokes beam, and the vibrational mode, respectively.

While the basic understanding of stimulated Raman processes can be found in literature, information is often scattered or embedded in disparate notations. Here key steps are elaborated for the sake of completeness and self-containment. We start with the field interaction of light with molecules. In SRS, the molecule is under the simultaneous interaction of two incident electric fields:

$$\boldsymbol{E}_{\omega_p} = E_p \boldsymbol{e_p} \sin \omega_p t \qquad (2a)$$

$$\boldsymbol{E}_{\omega_S} = E_S \boldsymbol{e_S} \sin \omega_S t \qquad (2b)$$

where $\omega_p$ and $\omega_S$ are the frequencies of the pump and Stokes laser beam fields, $E_p$ and $E_S$ are the corresponding amplitude, $\mathbf{e_p}$ and $\mathbf{e_S}$ are the unit vector that points to the polarization direction of the fields. In the electric dipole approximation of the molecule, the induced dipole moment $\boldsymbol{\mu}$ by the pump field $\omega_p$ can be expressed as $\boldsymbol{\mu}^{\text{ind}} = \alpha^{(3)} \cdot \boldsymbol{E}_{\omega_p}$ where $\alpha^{(3)}$ is the rank two 3×3 polarizability tensor of the molecule. Since $\boldsymbol{E}_{\omega_p}$ is an oscillatory function of time, then so will be $\boldsymbol{\mu}^{\text{ind}}$. Classically this oscillating dipole acts as an antenna and radiates. The interaction of this induced dipole with the Stokes field $\boldsymbol{E}_{\omega_S}$ gives rise to an energy as

$$V(t) = -\boldsymbol{E}_{\omega_S}(t) \cdot \boldsymbol{\mu}^{\text{ind}} = -\boldsymbol{E}_{\omega_S}(t) \cdot \alpha^{(3)} \cdot \boldsymbol{E}_{\omega_p}(t) \tag{3}$$

which appears in the form of a second-order nonlinear interaction. If we plug in Eq. (2) for the expression of two electric fields, after multiplying we get:

$$V(t) = \frac{1}{4} E_p E_S \cdot (\boldsymbol{e_S} \cdot \alpha^{(3)} \cdot \boldsymbol{e_p}) \cdot \left[ e^{i(\omega_p+\omega_S)t} + e^{-i(\omega_p+\omega_S)t} - e^{i(\omega_p-\omega_S)t} - e^{-i(\omega_p-\omega_S)t} \right] \tag{4}$$

This time-dependent interaction can induce four classes of transitions[23]. $e^{i(\omega_p+\omega_S)t}$ and $e^{-i(\omega_p+\omega_S)t}$ terms correspond to two-photon emission and two-photon absorption, respectively. Suppose $\omega_p > \omega_S$, then $e^{-i(\omega_p-\omega_S)t}$ relates to a process with an absorption emission of a pump photon and emission of a Stokes photon, i.e. a Stokes Raman scattering process. $e^{i(\omega_p-\omega_S)t}$ is then related to the anti-Stokes Raman scattering process.

For SRS in the form of pump loss and Stokes gain, the related interaction term is

$$V_{\text{SRS}}(t) = \frac{1}{4} E_p E_S \cdot (\boldsymbol{e_S} \cdot \alpha^{(3)} \cdot \boldsymbol{e_p}) \cdot e^{-i(\omega_p-\omega_S)t} \tag{5}$$

Now we can apply Fermi's golden rule derived from the time-dependent perturbation theory. The general form of Fermi's golden rule in the delta function representation is

$$w_{fi}(E) = \frac{2\pi}{\hbar} |V_{fi}|^2 \delta(E_f - E_i) \tag{6a}$$

where $V_{fi}$ is the matrix element of the coupling between the initial state $i$ to the final state $f$, and the Dirac delta function imposes conservation of energy between the initial and final state of the transition. Eq. (6a) can also be expressed in angular frequency instead of energy:

$$w_{fi}(\Omega_v) = \frac{2\pi}{\hbar^2} |V_{fi}|^2 \delta(\omega_S + \Omega_v - \omega_p) \tag{6b}$$

where we use $\Omega_v$ to denote the molecular intrinsic frequency of Raman mode. The energy conservation is ensured by the frequency restriction between the incident pump and Stokes laser beams and the frequency of the excited Raman mode. In other words, only vibrational state satisfying $\omega_S + \Omega_v - \omega_p = 0$ can be reached. Note that the transition probability $w_{fi}$ has a unit of $s^{-1}$, thus representing the probability of transition per unit time (i.e. rate $R_{\text{SRS}}$).

If we plug Eq. (5) into Eq. (6b), then we have the form of the golden rule in SRS transition:

$$R_{\text{SRS}} = \frac{\pi}{8\hbar^2} \cdot E_p^2 E_S^2 \cdot \left| \boldsymbol{e_S} \cdot \alpha^{(3)} \cdot \boldsymbol{e_p} \right|^2 \cdot \delta(\omega_S + \Omega_v - \omega_p) \qquad (7)$$

In realistic situations, the intrinsic Raman mode of the molecule does not have a perfectly well-defined transition frequency but is always spread into a continuous distribution by various broadening mechanisms. One often expresses this effect by stating that the final state is spread into a density of final state continuum[24]. In the current context of Raman scattering, this density of state is essentially the normalized lineshape profile $G(\Omega_v)$:

$$\int_0^\infty G(\Omega_v) \cdot d\Omega_v = 1 \qquad (8)$$

Then for a transition characterized by a density of final states, the final rate must be averaged over all possible values of the transition frequency, i.e., via integration over the lineshape profile.

$$R_{\text{SRS}} = \frac{\pi}{8\hbar^2} \cdot E_p^2 E_S^2 \cdot \left| \boldsymbol{e_S} \cdot \alpha^{(3)} \cdot \boldsymbol{e_p} \right|^2 \cdot \int_0^\infty G(\Omega_v) \cdot \delta(\omega_S + \Omega_v - \omega_p) \cdot d\Omega_v \qquad (9)$$

Then, one has

$$R_{\text{SRS}} = \frac{\pi}{8\hbar^2} \cdot E_p^2 E_S^2 \cdot \left| \boldsymbol{e_S} \cdot \alpha^{(3)} \cdot \boldsymbol{e_p} \right|^2 \cdot G(\Omega_v = \omega_p - \omega_S) \qquad (10)$$

where the notation $G(\Omega_v = \omega_p - \omega_S)$ means that the lineshape profile is to be evaluated at the frequency difference, $\omega_p - \omega_S$, of the incident pump and Stokes laser beams. Note that this notation applies to any frequency difference of the incident beams (not necessarily targeting at the peak of the Raman band), provided the lineshape function is known either experimentally or through a phenomenological function such as a Lorentzian profile.

In classical electromagnetics, the square of the field amplitude is related to the light intensity $I$ through $I = \frac{1}{2} c \varepsilon_0 E^2$ where $\varepsilon_0$ is the vacuum permeability constant[24]. Then Eq. (10) becomes

$$R_{\text{SRS}} = \frac{\pi}{2\varepsilon_0^2 c^2 \hbar^2} \cdot I_p I_S \cdot \left| \boldsymbol{e_S} \cdot \alpha^{(3)} \cdot \boldsymbol{e_p} \right|^2 \cdot G(\Omega_v = \omega_p - \omega_S) \qquad (11\text{a})$$

Further converting the light intensity to the photon flux, via $\phi = I/(\hbar \omega)$, one has

$$R_{\text{SRS}} = \frac{\pi \omega_p \omega_S}{2\varepsilon_0^2 c^2} \cdot \phi_p \phi_S \cdot \left| \boldsymbol{e_S} \cdot \alpha^{(3)} \cdot \boldsymbol{e_p} \right|^2 \cdot G(\Omega_v = \omega_p - \omega_S) \qquad (11\text{b})$$

In most experiments with gases, liquids and biomaterials, Raman modes are randomly orientated. Hence an additional factor of 1/9 is added to compensate for the dipole-field alignment of the randomly oriented molecules. This arises from the second-order light-molecule interaction, as each interaction contributes to a factor of 1/3 as in the case of linear interaction. Subsequently we have

$$R_{\text{SRS}} = \frac{\pi \omega_p \omega_S}{18 \varepsilon_0^2 c^2} \cdot \phi_p \phi_S \cdot \left| \boldsymbol{e_S} \cdot \alpha^{(3)} \cdot \boldsymbol{e_p} \right|^2 \cdot G(\Omega_v = \omega_p - \omega_S) \qquad (12)$$

Finally, comparing the definitions of Eq. (1) and Eq. (12), we have

$$\sigma_{SRS}(\Omega_v) = \frac{\pi \omega_p \omega_S}{18\varepsilon_0^2 c^2} \cdot \left| \boldsymbol{e_S} \cdot \alpha^{(3)} \cdot \boldsymbol{e_p} \right|^2 \cdot G(\Omega_v = \omega_p - \omega_S) \qquad (13)$$

This is the semiclassical theoretical expression for $\sigma_{SRS}$ as a function of $\Omega_v$. It directly links the experimentally-determined cross section with the intrinsic properties of the molecule, i.e. the polarizability tensor $\alpha^{(3)}$. A common way to compute $\alpha^{(3)}$ is through the famous Kramers-Heisenberg dispersion formula[23]:

$$\left(\alpha_{fi}\right)_{xy} = \sum_n \left( \frac{\langle f|\mu_x|n\rangle \langle i|\mu_y|n\rangle}{\hbar\omega + \hbar\omega_{ni}} - \frac{\langle n|\mu_x|i\rangle \langle f|\mu_y|n\rangle}{\hbar\omega - \hbar\omega_{nf}} \right) \qquad (14)$$

where $\mu$ is the dipole moment operator.

The lineshape function $G(\Omega_v)$ of Raman spectral peaks in realistic samples can often be modeled with a Lorentzian profile $\mathcal{L}(\tilde{v}) = \frac{1}{\pi} \frac{\frac{1}{2}\Gamma}{(\tilde{v}-\tilde{v}_0)^2 + \left(\frac{1}{2}\Gamma\right)^2}$, with $\Gamma$ being its full-width-at-half-maximum (FWHM). Typically in an SRS experiment, the pump and Stokes beam are tuned to match the peak position, $\Omega_0$, of the Raman mode, i.e., $\omega_p - \omega_S = \Omega_0$. $G(\Omega_v = \Omega_0)$ evaluated at the peak of the Lorentzian profile is $\mathcal{L}(\tilde{v})|_{\tilde{v}=\tilde{v}_0} = \frac{2}{\pi \Gamma}$. At this peak position,

$$\sigma_{SRS}(\Omega_v = \Omega_0) = \frac{\omega_p \omega_S}{9\varepsilon_0^2 c^2 \Gamma} \cdot \left| \boldsymbol{e_S} \cdot \alpha^{(3)} \cdot \boldsymbol{e_p} \right|^2 \qquad (15)$$

**Full QED derivation of stimulated Raman cross section**

Until now we have assumed the number of photons is large enough to apply the classical electromagnetic theory where the light is treated as classical functions of coordinates and time. Here we derive stimulated Raman cross section using QED theory in which both fields and matter are quantum mechanical. We start with Fermi's golden rule applied to SRS:

$$R(\Omega_v) = \frac{2\pi}{\hbar^2} |M_{fi}|^2 \delta(\omega_S + \Omega_v - \omega_p) \qquad (16)$$

where R is the transition probability, $M_{fi}$ is the matrix element of the interaction. From a QED perspective, this interaction results in a transition of the radiation field from an initial state of $|n_p(\boldsymbol{k_p}), n_S(\boldsymbol{k_S})\rangle$ to a final state of $|(n_p-1)(\boldsymbol{k_p}), (n_S+1)(\boldsymbol{k_S})\rangle$, where $n_p$ and $n_S$ are the number of photons in the pump beam mode and the Stokes beam mode, respectively, and $\boldsymbol{k_p}$ and $\boldsymbol{k_S}$ are the wave vector (note that the polarization information is also included to keep the notation simpler) corresponding to frequency $\omega_p$ and $\omega_S$, respectively. In general, second-order perturbation theory can calculate the transition matrix element as

$$M_{fi} = \sum_n \frac{\langle f|\hat{H}_{ED}|n\rangle \langle n|\hat{H}_{ED}|i\rangle}{E_i - E_n} \qquad (17)$$

where $\hat{H}_{ED}$ is the electric dipole interaction Hamiltonian which, in the long wavelength approximation, keeps the dominant contribution from the expansion of the electric potential energy and neglects the magnetic and the high-order nonlinear terms[2]. The $n$ summation runs over all the intermediate virtual states, and the energy $E_n$ in the denominator includes contribution from both the molecule and the field.

$\hat{H}_{ED}$ takes a form as $\hat{H}_{ED} = -\hat{\boldsymbol{\mu}} \cdot \hat{\boldsymbol{E}}_T$ where $\hat{\boldsymbol{\mu}}$ is the dipole moment operator and $\hat{\boldsymbol{E}}_T$ is the operator of the transverse electric field:

$$\hat{\boldsymbol{E}}_T(\boldsymbol{r}) = i \sum_k \sqrt{\frac{\hbar \omega_k}{2\varepsilon_0 V}} \cdot \boldsymbol{e}_k \cdot \left( \hat{a}_k e^{i\boldsymbol{k} \cdot \boldsymbol{r}} - \hat{a}_k^\dagger e^{-i\boldsymbol{k} \cdot \boldsymbol{r}} \right) \qquad (18)$$

where $\hat{a}_k$ and $\hat{a}_k^\dagger$ are the annihilation and creation operators, respectively, and $\boldsymbol{e}$ designates the unit polarization vector, same as before[2]. They exhibit the remarkable property of destroying or creating a quanta of energy, a photon in QED, as manifested by the simple structures of the only non-vanishing matrix elements:

$$\langle n-1|\hat{a}|n\rangle = \sqrt{n} \quad \text{and} \quad \langle n+1|\hat{a}^\dagger|n\rangle = \sqrt{n+1} \qquad (19)$$

The interaction Hamiltonian can take both the $\hat{\boldsymbol{\mu}} \cdot \hat{\boldsymbol{E}}_T$ form adopted here and the other $\hat{\boldsymbol{p}} \cdot \hat{\boldsymbol{A}}$ form involving the operator of the vector potential $\hat{\boldsymbol{A}}$, which produces the same results. The full quantum $\hat{H}_{ED} = -\hat{\boldsymbol{\mu}} \cdot \hat{\boldsymbol{E}}_T$ is analogous to the semiclassical case of $V = -\boldsymbol{\mu} \cdot \boldsymbol{E}$ used in Eq. (3) above.

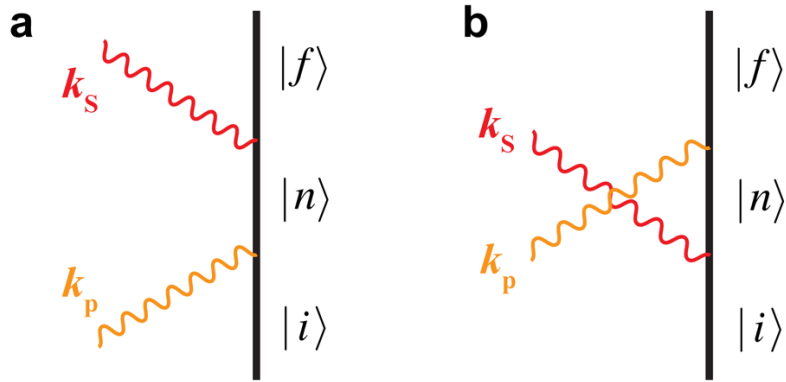

**Figure 2 Feynman diagram of stimulated Raman scattering.** Panel (**a**) shows the direct pathway where a molecule initially in the ground state $|i\rangle$ absorbs a pump photon with wavevector $\boldsymbol{k}_p$ and transitions to a virtual state $|n\rangle$. Subsequently, the molecule emits a Stokes photon $\boldsymbol{k}_S$ and transitions to the final state $|f\rangle$. Panel (**b**) depicts the inverse process, where the molecule first emits a Stokes photon before absorbing the pump photon. Both processes contribute to the net Raman scattering effect.

As depicted in **Figure 2**, there are two time-ordered pathways that can contribute to the matrix elements in SRS process. The first is where a molecule is transitioning from the initial state $i$ to a virtual intermediate state $n$ with the incident photon destroyed from the pump beam, followed by the emission of a new photon to the Stokes beam and molecular transition to the final state *f*. The second is similar, with the photon creation "preceding" the photon destruction.

All these formulas and considerations above allow explicit evaluation of the matrix element in Eq. (17) as:

$$M_{fi} = \sqrt{\frac{n_p \hbar \omega_p}{2\varepsilon_0 V}} \sqrt{\frac{(n_S+1)\hbar \omega_S}{2\varepsilon_0 V}} \left[ \sum_n \left\{ \frac{(\mu^{fn} \cdot e_S)(\mu^{ni} \cdot e_p)}{E_n - E_i - \hbar \omega_p - i\hbar\gamma} + \frac{(\mu^{fn} \cdot e_p)(\mu^{ni} \cdot e_S)}{E_n - E_i + \hbar \omega_S - i\hbar\gamma} \right\} \right] \quad (20a)$$

where $\mu$ is the dipole moment vector. The first term in the summation with energy difference in the denominator corresponds to the diagram (a), and the second term with energy sum in the denominator corresponds to the diagram (b). $i\hbar\gamma$ is included as a damping term. Clearly, $\sqrt{n_p}$ arises as a consequence of $\hat{a}_k$ acting on the transition from $|n_p(k_p)\rangle$ to $|(n_p-1)(k_p)\rangle$; $\sqrt{n_S + 1}$ arises as a consequence of $\hat{a}_k^\dagger$ acting on the transition from $|n_S(k_S)\rangle$ to $|(n_S+1)(k_S)\rangle$. For simplicity, we use $\mathbb{M}$ to denote the summation part:

$$\mathbb{M} \equiv \sum_n \left\{ \frac{(\mu^{fn} \cdot e_S)(\mu^{ni} \cdot e_p)}{E_n - E_i - \hbar \omega_p - i\hbar\gamma} + \frac{(\mu^{fn} \cdot e_p)(\mu^{ni} \cdot e_S)}{E_n - E_i + \hbar \omega_S - i\hbar\gamma} \right\} \quad (20b)$$

Substituting the matrix element of Eq. (20) to Eq. (16) yields the rate:

$$R_{SRS} = \frac{\pi \omega_p \omega_S}{2\varepsilon_0^2 c^2} \cdot \left(\frac{n_p c}{V}\right) \left(\frac{(n_S+1) c}{V}\right) \cdot |\mathbb{M}|^2 \cdot G(\Omega_v = \omega_p - \omega_S) \quad (21)$$

where we have also introduced the lineshape profile and integrated it over the Dirac delta function as we did in Eq. (9).

The incident photon flux $\phi$ is defined as the number of photons per unit time per unit area that crosses a given point. If the volume *V* consists of a photon beam of area *A* and length *L*, then in a time *t=L/c* the number of the photons crossing is simply the number of photons, n, within this volume. Thus one can easily verify that

$$\phi = \frac{n}{A \cdot t} = \frac{n}{A \cdot (L/c)} = \frac{n \cdot c}{V} \quad (22)$$

Obviously, $n_S + 1 \approx n_S$ for strong laser beams employed in real SRS experiments. Introducing the photon flux to Eq. (21), and again adding the orientation averaged factor of 1/9 yields

$$R_{SRS} = \frac{\pi \omega_p \omega_S}{18 \varepsilon_0^2 c^2} \cdot \phi_p \phi_S \cdot |\mathbb{M}|^2 \cdot G(\Omega_v = \omega_p - \omega_S) \quad (23)$$

Comparing the rate equation Eq. (23) to the definition of Eq. (1), we arrive at the expression for stimulated Raman cross section:

$$\sigma_{SRS}(\Omega_v) = \frac{\pi \omega_p \omega_S}{18 \varepsilon_0^2 c^2} \cdot |\mathbb{M}|^2 \cdot G(\Omega_v = \omega_p - \omega_S) \quad (24a)$$

This is the QED expression for $\sigma_{SRS}$. To our knowledge, this is the first time a rigorous expression is formally given for $\sigma_{SRS}$. Its structure is nearly identical to that of Eq. (13), derived from a semiclassical theory. Again, when the pump and Stokes beam are tuned to match the peak position, $\Omega_0$, of the Raman mode (assumed with a Lorentzian profile), one has a simple form for its peak value

$$\sigma_{SRS}(\Omega_v = \Omega_0) = \frac{\omega_p \omega_S}{9\varepsilon_0^2 c^2 \Gamma} \cdot |\mathbb{M}|^2 \tag{24b}$$

It is insightful to relate this to a common but often mysterious dimensionless factor -- the fine structure constant $\alpha$ in fundamental physics:

$$\alpha = \frac{1}{4\pi\varepsilon_0} \frac{e^2}{\hbar c} \approx \frac{1}{137} \tag{25}$$

To display the terms contained in the constant, we extract the elementary charge $e$ and reduced Planck constant $\hbar$ from $\mathbb{M}$ in Eq. (24a):

$$\sigma_{SRS} = \frac{\pi e^4 \omega_p \omega_S}{18\varepsilon_0^2 \hbar^2 c^2} \cdot \left|\sum_n \left\{\frac{(\boldsymbol{D}^{fn}\cdot\boldsymbol{e}_S)(\boldsymbol{D}^{ni}\cdot\boldsymbol{e}_p)}{\omega_n - \omega_i - \omega_p - i\gamma} + \frac{(\boldsymbol{D}^{fn}\cdot\boldsymbol{e}_p)(\boldsymbol{D}^{ni}\cdot\boldsymbol{e}_S)}{\omega_n - \omega_i + \omega_S - i\gamma}\right\}\right|^2 \cdot G(\Omega_v) \tag{26}$$

where the dipole moment $\boldsymbol{\mu}$ is reduced to $\boldsymbol{D}$, the displacement vector. Then it is easy to convert Eq. (26) into

$$\sigma_{SRS} = \frac{8\pi^3}{9}\alpha^2 \cdot \omega_p \omega_S \cdot |\overline{\mathbb{M}}|^2 \cdot G(\Omega_v) \tag{27a}$$

where we have re-defined the matrix element as

$$\overline{\mathbb{M}} = \sum_n \left\{\frac{(\boldsymbol{D}^{fn}\cdot\boldsymbol{e}_S)(\boldsymbol{D}^{ni}\cdot\boldsymbol{e}_p)}{\omega_n - \omega_i - \omega_p - i\gamma} + \frac{(\boldsymbol{D}^{fn}\cdot\boldsymbol{e}_p)(\boldsymbol{D}^{ni}\cdot\boldsymbol{e}_S)}{\omega_n - \omega_i + \omega_S - i\gamma}\right\} \tag{27b}$$

This compact form of Eq. (27), together with the dimensionless nature of α, prompts us to readily verify the unit of $\sigma_{SRS}$. The numerator of $\overline{\mathbb{M}}$ apparently carries a dimension of length squared. The denominator of $\overline{\mathbb{M}}$ carries a dimension of $\omega^2$, which cancels out with that of $\omega_p\omega_S$. The remaining $G(\Omega_v=\omega_p-\omega_S)$ exhibits a dimension of time (sec). Thus, the unit of the overall expression turns out to be $m^4 \cdot s$, as expected in Göppert-Mayer (1 GM = $10^{-50}$ $cm^4 \cdot s$), the unit introduced for two-photon absorption cross sections, $\sigma_{TPA}$[24]. This dimensional analysis also sheds light on the physical meaning of the factors that determine $\sigma_{SRS}$: the $\frac{8\pi^3}{9}$ factor correlates to the polarization orientation and spatial angle, $\alpha^2$ denotes the probability of a second-order field-matter interaction (as $\alpha$ governs the strength of electromagnetic interaction between charged particles and photons), $G(\Omega_v)$ captures the time scale, and $\omega_p\omega_S \cdot |\overline{\mathbb{M}}|^2$ together determines the spatial scale.

## Numerical estimation and comparison with experiments

Numerical evaluation can be facilitated by Eq. (27) expressed with the fine structure constant α. First we might be able to estimate the order of magnitude of $\sigma_{SRS}$ for Raman modes in small molecules far away from electronic resonance. Strictly speaking, the exact evaluation requires a sum-over-state calculation, as the number of electronic states contributing to the polarizability is large. However, a back-of-the-envelope approximation is useful too. If we excite small molecules whole electronic state lies in the UV around 200 nm by a laser excitation around 1000 nm, then the angular frequency dependence in $\omega_p \omega_S \cdot |\overline{M}|^2$ will produce a value close to $\left(\frac{1/1000}{1/200}\right)^2 \approx 0.04$ where the damping term can be neglected in far off resonance. We then assume the displacement $D$ of the transition dipole moment to be 1/10 of the length of a bond (assumed to be around 1.5 Å) and the linewidth $\Gamma$ of a typical Raman mode in condensed phase to be 15 cm$^{-1}$, equivalent to $4.5 \times 10^{11}$ rad·s$^{-1}$. Then Eq. (27) predicts

$$\sigma_{SRS} \approx \frac{8\pi^3}{9} \times \left(\frac{1}{137}\right)^2 \times 0.04 \times (1.5 \times 10^{-11} \text{ m})^4 \times \frac{2}{\pi \times 4.5 \times 10^{11} \text{ rad·s}^{-1}} = 4 \times 10^{-2} \text{ GM} \quad (28)$$

This estimated result of 0.04 GM corresponds well with the experimentally measured value of 0.04 GM from C-O bond of small molecule methanol[20]. While this is not meant to be a rigorous calculation, the agreement between theory and experiment is encouraging.

Electronic resonance can drastically enhance the cross sections of electronically coupled Raman modes, as the detuning approaches zero in the denominator. The summation over electronic states can be relaxed, provided the single resonant state has a large enough transition dipole moment. By applying the Born-Oppenheimer approximation of separability of electronic and vibrational wavefunction, resonance Raman has been shown to be dominated by the so-called Albrecht's A-term for strongly allowed electronic transitions and substantial nonorthogonality of Frank-Codon overlap factor[25,26]. Let's assume both the laser excitation wavelength and the chromophore electronic absorption to be around 700 nm (i.e., exact electronic resonance) which is about 14,000 cm$^{-1}$, and the electronic linewidth (damping term) to be around 700 cm$^{-1}$, the frequency dependence of $\omega_p \omega_S \cdot |\overline{M}|^2$ will produce $\left(\frac{14000 \text{ cm}^{-1}}{700 \text{ cm}^{-1}}\right)^2 = 400$. Note this factor is 10,000 folds higher than the far-off resonance case above. Electronic resonance also creates strong displacement $D$ for electronically coupled Raman modes. For example, electronic transition dipole moments have been reported as 7.1 Debye (about 1.4 $e$·Å) for PM546 dye and 8.1 Debye (about 1.6 $e$·Å) for Rhodamine 123, respectively[27]. If we take the transition displacement $D$ as 1.6Å, the numerator of $|\overline{M}|^2$ will produce another factor of 13,000 compared to the small molecule above. Finally we need to consider Frank-Codon overlap when evaluating the dipole moment elements of vibronic transitions. For strongly coupled transition, this overlap can be substantial according to Albrecht's theory if there is sizable shift of the excited state potential along the vibrational coordinate[25,26]. Without loss of generality, it is assumed to 0.1 here. Together, Eq. (27) predicts the electronic resonant result to be

$$\sigma_{SRS} \approx \frac{8\pi^3}{9} \times \left(\frac{1}{137}\right)^2 \times 400 \times (1.6 \times 10^{-10} \text{ m})^4 \times \frac{2}{\pi \times 4.5 \times 10^{11} \text{ rad·s}^{-1}} \times 0.1 \approx 5 \times 10^5 \text{ GM} \quad (29)$$

Experimentally, in the exact electronic resource condition, R6G has a stimulated Raman cross section of 860,000 GM for the electronic coupled ring mode, and, similarly, IR820 chromophore has around 430,000 GM[20]. Again, the agreement is satisfactory considering the crude approximation. The success of both the off resonance case in Eq. (28) and the electronic resonance case in Eq. (29) indicates that the theory presented here can predict outcome of SRS experiments from first principles.

## Intrinsically strong Raman response: comparison with two-photon absorption cross sections

TPA cross sections, $\sigma_{TPA}$, have been well documented in the literature, thus serving as a natural reference for nonlinear light-molecule interaction. A somewhat surprising finding is that $\sigma_{SRS}$ compare rather favorably to that $\sigma_{TPA}$[20], in stark contrast to the prevailing perspective about the comparison between $\sigma_{Raman}$ (vibrational) and $\sigma_{absorption}$ (electronic) in which the former is more than 10 orders of magnitude smaller. There are two manifestations. First, $\sigma_{SRS}$ from small molecules are generally not too much smaller than $\sigma_{TPA}$ of common chromophores. For example, standard dyes such as fluorescein and eGFP display $\sigma_{TPA}$ in the vicinity of 100 GM. Small molecules containing one or two C≡C modes, such as EdU and conjugated 2-yne, have $\sigma_{SRS}$ in the comparable range of 5-100 GM. Stimulated Raman response is not that weak after all, considering the smaller size of the vibrational moiety. Second, $\sigma_{SRS}$ can even surpass $\sigma_{TPA}$ for molecules that are experiencing electronic resonance. For example, R6G exhibits a near record-high $\sigma_{SRS}$ around 860,000 GM while its $\sigma_{TPA}$ is only around 100 GM based on various reports. In comparison, $\sigma_{TPA}$ on the level of 100,000 GM has not been reported in the literature for small organic molecules, to the best of our knowledge.

Our theoretical expression can provide valuable insights and semi-quantitative explanation towards this observation. One factor that works in favor of $\sigma_{SRS}$ is the factor of $G(\Omega_v=\omega_p-\omega_S)$. In the semiclassical theory $e^{-i(\omega_p+\omega_S)t}$ term in Eq. (4) is responsible for two-photon absorption, and the final expression of $\sigma_{TPA}$ also contains its own lineshape profile[24]. In condensed phase, Raman bands are usually ~100 times narrower (i.e., smaller values of $\Gamma$) than the line profile of TPA of chromophores[8,28], resulting in a significantly higher density of state and hence the value of $G(\Omega_v=\Omega_0)$. This difference of ~100 is a large and general factor that applies to nearly all comparison between $\sigma_{SRS}$ and $\sigma_{TPA}$. It is likely a key reason as to why $\sigma_{SRS}$ from small molecules are not too much smaller than $\sigma_{TPA}$ of common chromophores.

Another factor is the favorable transition dipole moments during the electronic resonance. The resonant SRS response involves the transition dipole moment between the ground state ($S_0$) and the electronic excited state ($S_1$) four times. In contrast, a resonant TPA response involves the transition dipole moment between $S_0$ and $S_1$ two times and that between $S_1$ and $S_n$ two times, an overall four-field interaction too. For strongly absorbing dyes such as R6G, their ground-state molecular extinction coefficients are at the largest possible values empirically – one can hardly find another molecule with much larger ground-state molecular extinction coefficient. Hence, the corresponding transition dipole moments between $S_0$ and $S_1$ shall approach the physical maximum for molecules of their sizes. In comparison, the transition dipole moments between $S_1$ and $S_n$ shall be weaker or comparable at most, which indeed is the case[20,29]. Hence, $\sigma_{SRS}$ can benefit from the large transition dipole moments two more times when compared to $\sigma_{TPA}$. Together, the sharp vibrational lineshape and the favorable resonance enhancement when approaching electronic resonance both contribute to the relatively strong $\sigma_{SRS}$.

## Spontaneous Raman cross sections

Quantum mechanical expression for σ$_{Raman}$ has been given in the literature[2,26,30]. Yet it will be constructive to derive it under the aforementioned notation, as some intermediate steps will be needed for subsequent comparison with σ$_{SRS}$. We start the semiclassical treatment with Fermi's Golden rule similar to Eq. (7):

$$R_{fi} = \frac{\pi}{8\hbar^2} \cdot E_p^2 E_S^2 \cdot |\boldsymbol{e_S} \cdot \alpha^{(3)} \cdot \boldsymbol{e_p}|^2 \cdot \delta(\omega_f - \omega_i) \tag{30}$$

In spontaneous Raman, all vacuum modes are accessible for the scattered photon, and thus we can take the sum of different modes with different wavevectors $\boldsymbol{k}$:

$$R_{Raman} = \frac{\pi}{2\varepsilon_0^2 c^2 \hbar^2} \cdot \Sigma_{\boldsymbol{k_S}} I_p I_S \cdot |\boldsymbol{e_S} \cdot \alpha^{(3)} \cdot \boldsymbol{e_p}|^2 \cdot \delta(\omega_S + \Omega_v - \omega_p) \tag{31}$$

where we have also used $I = \frac{1}{2}\varepsilon_0 c E^2$. $I_p$ is directly related to the photon flux $\phi_p$ via $I_p = \phi_p \cdot \hbar\omega_p$.

A key task here is how to treat the Stokes intensity $I_S$. In the strictly semiclassical theory in which light is treated as a classical electromagnetic field, spontaneous Raman scattering does not occur. According to Eq. (3), the absence of a classical Stokes field nullifies the strength of the nonlinear coupling between the molecular transition and the fields, so that the scattering rate vanishes. Hence we have to "borrow" the concept from QED and adopt a shortcut approach. Let's consider a small rectangular cuboid region (a volume $V$ defined by an area $A$ and a length $L$) within which the modes are defined. By definition, the light intensity equals the number of photons contained in this region, m$_{vacuum}$, multiplied by the energy per photon and divided by the cross-sectional area of the region and by the transit time through the region—that is,

$$I_S = \frac{m_{vacuum} \cdot \hbar \omega_S}{A \cdot (L/c)} = \frac{m_{vacuum} \cdot \hbar \omega_S}{V/c} \tag{32}$$

Borrowing the picture of QED, one can treat $I_S$ as an *effective* intensity contributed from vacuum fluctuation and consider each mode contains one *virtual* photon from vacuum zero-point fluctuations, i.e. $m_{vacuum}=1$ in Eq. (32). Therefore Eq. (31) becomes

$$R_{Raman} = \frac{\pi}{2\varepsilon_0^2 c^2 \hbar^2} \cdot \Sigma_{\boldsymbol{k_S}} (\phi_p \cdot \hbar\omega_p) \cdot \left(\frac{1 \cdot \hbar \omega_S}{V/c}\right) \cdot |\boldsymbol{e_S} \cdot \alpha^{(3)} \cdot \boldsymbol{e_p}|^2 \cdot \delta(\omega_S + \Omega_v - \omega_p) \tag{33}$$

The summation over all the scattered wavevectors can be converted into an integration over frequency $\omega_S$ and solid angle $\Omega$, a common practice used in electrodynamic theory[2]:

$$\Sigma_{\boldsymbol{k_S}} \to \frac{V}{(2\pi)^3} \iint d\omega_S d\Omega \frac{\omega_S^2}{c^3} \tag{34}$$

Then the rate in Eq. (33), now considered as a differential rate into the solid angle d$\Omega$, becomes

$$\frac{dR_{Raman}}{d\Omega} = \frac{1}{16\pi^2 \varepsilon_0^2 c^4} \cdot \int \omega_p \omega_S^3 \cdot \phi_p \cdot |\boldsymbol{e_S} \cdot \alpha^{(3)} \cdot \boldsymbol{e_p}|^2 \cdot \delta(\omega_S + \Omega_v - \omega_p) \cdot d\omega_S \tag{35}$$

A non-zero contribution to the total rate would require Raman resonance conditions at $\omega_S=\omega_p-\Omega_v$. The frequency integration is readily performed with the use of delta function:

$$\frac{dR_{Raman}}{d\Omega} = \frac{1}{16\pi^2\varepsilon_0^2 c^4} \cdot \omega_p \omega_S^3 \cdot \phi_p \cdot |\boldsymbol{e_S} \cdot \alpha^{(3)} \cdot \boldsymbol{e_p}|^2 \tag{36}$$

Integrating $d\Omega$ over all the spatial angel ($4\pi$) and assume isotropic scattering under two polarizations of light, and also adding the 1/9 factor for orientation, then the total Raman scattering rate can be evaluated at

$$R_{Raman} = \frac{1}{18\pi\varepsilon_0^2 c^4} \cdot \omega_p \omega_S^3 \cdot \phi_p \cdot |\boldsymbol{e_S} \cdot \alpha^{(3)} \cdot \boldsymbol{e_p}|^2 \tag{37}$$

If defined by flux $R_{Raman} = \sigma'_{Raman} \cdot \phi_p$, then we arrive at

$$\sigma'_{Raman} = \frac{\omega_p \omega_S^3}{18\pi\varepsilon_0^2 c^4} \cdot |\boldsymbol{e_S} \cdot \alpha^{(3)} \cdot \boldsymbol{e_p}|^2 \tag{38a}$$

If defined as $P_{Raman} = \sigma_{Raman} \cdot I_p$ where $P$ is the power scattered into the Stokes channel, then we arrive at

$$\sigma_{Raman} = \frac{\omega_S^4}{18\pi\varepsilon_0^2 c^4} \cdot |\boldsymbol{e_S} \cdot \alpha^{(3)} \cdot \boldsymbol{e_p}|^2 \tag{38b}$$

Note the additional factor of $\omega_S/\omega_p$ between these two different definitions of Raman cross sections. Eq. (38) is the semiclassical theory for $\sigma_{Raman}$, with a hybrid concept of virtual photons from vacuum contribution.

The full QED result can be naturally obtained by assigning $n_S=0$ in the matrix element of Eq. (20). Repeating the steps leading to Eq. (21) and taking the sum of different modes results in

$$R_{Raman} = \sum_{\boldsymbol{k_S}} \frac{\pi \omega_p \omega_S}{2\varepsilon_0^2 c^2} \cdot \left(\frac{n_p c}{V}\right)\left(\frac{c}{V}\right) \cdot |\mathbb{M}|^2 \cdot \delta(\omega_S + \Omega_v - \omega_p) \tag{39}$$

which has the same structure as Eq. (33) after replacing $\frac{n_p c}{V}$ by $\phi_p$. Repeating the subsequent procedure, we can obtain the final result

$$\sigma'_{Raman} = \frac{\omega_p \omega_S^3}{18\pi\varepsilon_0^2 c^4} \cdot |\mathbb{M}|^2 \tag{40a}$$

$$\sigma_{Raman} = \frac{\omega_S^4}{18\pi\varepsilon_0^2 c^4} \cdot |\mathbb{M}|^2 \tag{40b}$$

As it is transparent from the derivation above, the full QED considers the coupling with vacuum state ($n_S=0$) in a straightforward manner rather than in an *ad hoc* fashion as in the semiclassical treatment.

## Einstein-coefficient-like equation for Raman scattering and its generalization

We are finally in a position to investigate the connection between σ$_{Raman}$ and σ$_{SRS}$. Comparing Eq. (40b) with Eq. (24b), one can connect the peak value of $\sigma_{SRS}(\Omega_v = \Omega_0)$ to σ$_{Raman}$ as

$$\sigma_{Raman} = \frac{\omega_S^3 \Gamma}{2\pi c^2 \omega_p} \sigma_{SRS}(\Omega_v = \Omega_0) \tag{41}$$

This reproduces the exact equation that was recently derived using the concept of virtual vacuum photons but without explicitly referring to full quantum mechanics[21]. This relation carries the spirit of Einstein's A and B coefficients connecting the rate of spontaneous emission and stimulated emission. It has found utility in predicting and explaining absolute signal of SRS microscopy such as the enhancement factor and signal-to-noise ratio [22]. Now we have proved its validity in an independent way.

However, it might be misleading to perceive Eq. (41) as a mere proportionality relation. Its physical meaning becomes more transparent if expressed in an integral form. As shown in Eq. (24a), σ$_{SRS}$ itself is a function of $\Omega_v$. Then it is easy to verify the following equation regarding $\sigma'_{Raman}$ calculated in Eq. (38a):

$$\sigma'_{Raman} = \frac{\omega_S^2}{\pi^2 c^2} \int \sigma_{SRS}(\Omega_v) d\Omega_v \tag{42}$$

This explicitly states that Raman cross section is an integration of the stimulated Raman cross section instead of a simple proportionality relation. This is an important but subtle distinction in the nature of these two. It is also consistent with experimental procedures that experimentalists take to measure these two. Stimulated Raman can be excited by tuning the difference, $\omega_p$-$\omega_S$, of the incident pump and Stokes laser beams to target any position of the Raman lineshape profile, and the subsequent rate and $\Omega_v$-dependent cross section (in the unit of Göppert-Mayer) can be readily determined by the definition of Eq. (1). Further integration over the peak will yield a unit that is not Göppert-Mayer. This is exactly opposite in spontaneous Raman experiment, in which the entire Raman scattering peak has to be integrated to obtain the energy flux in order to report the Raman cross section (in the unit of cm$^2$) of the mode. Theoretically, the concept of "Raman lineshape" is not needed in modeling spontaneous Raman scattering, both quantum mechanically and pure classically. In contrast, both classical and quantum theories have to incorporate a vibrational damping term (Raman line broadening) to explain stimulated Raman scattering – otherwise the energy flow from the fields to the molecules would become unbounded (i.e., infinity).

The integral form of Eq. (42) also implies the existence of a differential form. Following the QED result of Eq. (39), one can integrate the d$\Omega$ and polarization but leave d$\omega_S$ as

$$\frac{dR_{Raman}}{d\omega_S} = \frac{1}{18\pi\varepsilon_0^2 c^4} \cdot \omega_p \omega_S^3 \cdot \phi_p \cdot |\mathbb{M}|^2 \cdot \delta(\omega_S + \Omega_v - \omega_p) \tag{43}$$

Similar to the treatment of Dirac delta function above in Eq. (9), we can define the lineshape profile and integrate it over

$$\frac{dR_{\text{Raman}}}{d\omega_S} = \frac{1}{18\pi\varepsilon_0^2 c^4} \cdot \omega_p \omega_S^3 \cdot \phi_p \cdot |\mathbb{M}|^2 \cdot G(\Omega_v = \omega_p - \omega_S) \tag{44}$$

Dividing $\phi_p$ from the differential rate defines a spectral-differential Raman cross section (in the unit of cm² per frequency) as

$$\frac{d\sigma'_{\text{Raman}}}{d\omega_S} = \frac{\omega_p \omega_S^3}{18\pi\varepsilon_0^2 c^4} \cdot |\mathbb{M}|^2 \cdot G(\Omega_v = \omega_p - \omega_S) \tag{45}$$

Indeed, comparing Eq. (45) with Eq. (24a) leads to the differential form:

$$\frac{d\sigma'_{\text{Raman}}}{d\omega_S} = \frac{\omega_S^2}{\pi^2 c^2} \sigma_{\text{SRS}}(\Omega_v) \tag{46}$$

which complements the integral form of Eq. (42). However, the $\frac{d\sigma'_{\text{Raman}}}{d\omega_S}$ is not commonly reported in the literature.

Finally it is constructive to compare our result to Füchtbauer-Ladenburg equation, which was introduced long time ago for treating fluorescence in the literature of atomic physics[31]:

$$\frac{1}{\tau_{\text{rad}}} = \frac{8\pi}{c^2} \int \nu^2 \sigma_{\text{em}}(\nu) d\nu \tag{47}$$

where $\sigma_{\text{em}}(\omega)$ is the stimulated emission cross section and $\tau_{\text{rad}}$ is the lifetime of the upper level. Füchtbauer-Ladenburg equation is often regarded as a generalized form of Einstein's coefficient which was derived in the context of broadband blackbody radiation. Approximating the integral by moving term $\nu^2$ out of the integral and converting to angular frequency, Füchtbauer-Ladenburg equation can be rewritten as

$$\frac{1}{\tau_{\text{rad}}} = \frac{\omega_S^2}{\pi^2 c^2} \int \sigma_{\text{em}}(\omega) d\omega \tag{48}$$

which has an identical structure to Eq. (42) for Raman cross sections. Clearly the resemblance suggests the deep connection between emission and scattering events.

Based on the analogous structures revealed above, it is intriguing to propose a symmetric equation that connects all four processes:

$$\frac{1/\tau_{rad}}{\int \sigma_{\text{em}}(\omega) d\omega} = \frac{\sigma'_{\text{Raman}}}{\int \sigma_{\text{SRS}}(\Omega_v) d\Omega_v} \tag{49}$$

This relation could be considered as a generalized version of Einstein's coefficient, especially in the context of cavity QED. Indeed, in cavity QED[32], the rate of spontaneous emission could be controlled depending on the boundary conditions of the surrounding vacuum field. The possible enhancement or inhibition of the spontaneous emission rate is known as the Purcell effect. Similar effects have been observed in Raman scattering. In 1993, Cairo *et al.* put a Raman medium ($C_6H_6$) in a cavity and observed that it is possible to enhance or inhibit spontaneous Raman scattering for a specific Raman line, just by spectral tuning of the cavity[33]. Since then, more similar observations have been made[34–36]. Therefore, when the surrounding vacuum mode is significantly altered, the original Einstein's A and B coefficients, as well as the Füchtbauer-Ladenburg equation, would break down, and so would our relation, Eq. (42), between $\sigma_{Raman}$ and $\sigma_{SRS}$. However, their respective ratios should still be equal to each other, as spontaneous emission and spontaneous Raman are fundamentally driven by the same vacuum environment.

## Conclusion

Stimulated Raman cross section $\sigma_{SRS}$ was recently defined and measured as a quantitative property to characterize how molecules respond under coherent Raman scattering. In this work, we have derived a full theoretical expression for $\sigma_{SRS}$ through both a semi-classical and a QED treatment. The final result is also linked to the fine-structure constant for physical rationale of the formula. We also conducted numerical evaluation to determine the value of $\sigma_{SRS}$ for common small molecules as well as molecules under electronic resonance. The calculation corresponds well with experiment results, and reveals the physics (such as the sharp vibrational lineshape and the favorable resonance enhancement) underlying the strong stimulated Raman response. Then, we independently re-derived the recently reported bridging equation between $\sigma_{SRS}$ and $\sigma_{Raman}$, and discussed its integral and differential forms. Finally, we examined the distinct nature between the two sets of cross sections, and how it can be integrated with the emission counterpart (i.e., Füchtbauer-Ladenburg equation) towards a generalization of Einstein's original A and B coefficients.

## Acknowledgement

We are thankful for helpful discussion with Eric Potma, Sunney Xie, Lu Wei, Fanghao Hu, Lixue Shi, Warren Warren, Ji-Xin Cheng, Ara Apkarian, Graham Fleming, Na Ji, Minbiao Ji, Naixin Qian and Xuemeng Li. We acknowledge support from the Air Force Office of Scientific Research (AFOSR) (Grant No. FA9550-21-1-0170), the National Institute of Health (Grant No. R01 EB029523), and Chan Zuckerberg Initiative (Dynamic Imaging 2023-321166).